\documentclass{elsarticle}
\usepackage{color,graphicx,amsmath,array,xfrac,hyperref}
\usepackage{hyperref}

\journal{International Journal of Adhesion and Adhesives}

\begin{document}

\begin{frontmatter}

\title{Consideration of the restriction of lateral contraction in the elastic behaviour of cohesive zone models}
\tnotetext[mytitlenote]{doi:10.1016/j.ijadhadh.2015.07.003}
\tnotetext[mytitlenote2]{\copyright 2015. This manuscript version is made available under the CC-BY-NC-ND 4.0 license \href{http://creativecommons.org/licenses/by-nc-nd/4.0/}{http://creativecommons.org/licenses/by-nc-nd/4.0/}}

\author{Olaf Hesebeck}
\address{Fraunhofer Institute for Manufacturing Technology and Advanced Materials, IFAM, Wiener Stra{\ss}e 12,
28359 Bremen, Germany}
\ead{olaf.hesebeck@ifam.fraunhofer.de}




\begin{abstract}
Cohesive zone models do not consider the lateral contraction of adhesive layers under tensile loads. The constraint of the lateral contraction by the adherents which depends on the geometry of the adhesive layer has a major influence on the normal stiffness of the joint. Two methods to improve the accuracy of the stiffness of cohesive zone models of rectangular adhesive layers are proposed in this paper. Both approaches use existing closed-form solutions for rectangular elastic layers between rigid plates. The first assigns an effective stiffness to the entire cohesive zone, the second approach defines a spatially varying stiffness to account for the difference in constraint of the adhesive close to the free surfaces and in the centre of the layer. The accuracy in joint stiffness for cohesive zone models gained by the two methods is tested in two extensive parametric studies considering both rigid and flexible adherents.
\end{abstract}

\begin{keyword}
Cohesive zone modelling \sep Elasticity \sep Finite element analysis \sep Constraint effects \sep Parameter identification
\end{keyword}

\end{frontmatter}


\section{Introduction}
Cohesive zone models based on traction-separation laws have widely been used to describe crack growth processes at interfaces. Their implementation in finite element codes has enabled several applications like the simulation of delamination in composites or the failure of adhesive joints. The development of methods to identify traction-separation relations from test results has been a focus of many research efforts in recent years.

The most important properties characterizing a traction-separation law in a single mode are the strength and the critical energy release rate. The influence of further details of the shape of the traction-separation curve on simulation results is in general much smaller. Sensitivity analysis of tests used for parameter identification like DCB, ENF, and SLJ concerning cohesive law parameters can be found e.g. in \cite{Alfano2006723,Gustafson20092201,Valoroso20101666}. An approach to a direct identification of the traction-separation law using the J-integral is applied e.g. in \cite{Stigh2000297,Sorensen20021053,Marzi2011840,Dias2013646,Marzi2014324}.

While the effect of the damage evolution law on the results of adhesive joint simulations using cohesive zone models has been considered by several authors, the choice of the elastic behaviour prior to damage initiation was not in the focus of research yet. 
In case of the modelling of delaminations in composites, the elastic stiffness is usually considered as a penalty parameter without physical meaning. The elasticity of the adhesive layer in an adhesive joint is meaningful, but the fracture mechanical tests for parameter identification are not very sensitive to the elastic parameters \cite{Valoroso20101666}. A negligible influence of the elastic properties was also observed in the simulation of a steel sheet structure under crash loads which was bonded using thin layers of an epoxy adhesive \cite{May2015112}. Possibly the effect of adhesive compliance is larger in case of flexible adhesive joints.

While the elastic properties of the cohesive zone models seem to be of minor importance for the simulation of the failure of adhesively bonded structures, it is still desirable from a practical point of view that the model exhibits the correct stiffness. If the finite element model of the adhesive layer give a sufficiently accurate stiffness, then the same model can be used as well for a static analysis as for a crash simulation, for example.

This article considers the modelling of the stiffness of adhesive joints using cohesive zone models. The results may be transferable to other applications of cohesive zone models where stiffness is of interest. Many cohesive laws assume a linear, uncoupled elastic behaviour which can be characterized by one stiffness parameter for the normal direction and another parameter describing the shear stiffness of the adhesive layer. A simple choice for the normal stiffness is the adhesive's elastic modulus divided by the adhesive layer thickness and for the shear stiffness the shear modulus divided by the thickness. While this is a good choice for the shear stiffness, the normal stiffness is in reality larger due to the restricted lateral contraction.

\begin{figure}[hbtp]
\centering
\includegraphics[width=8cm]{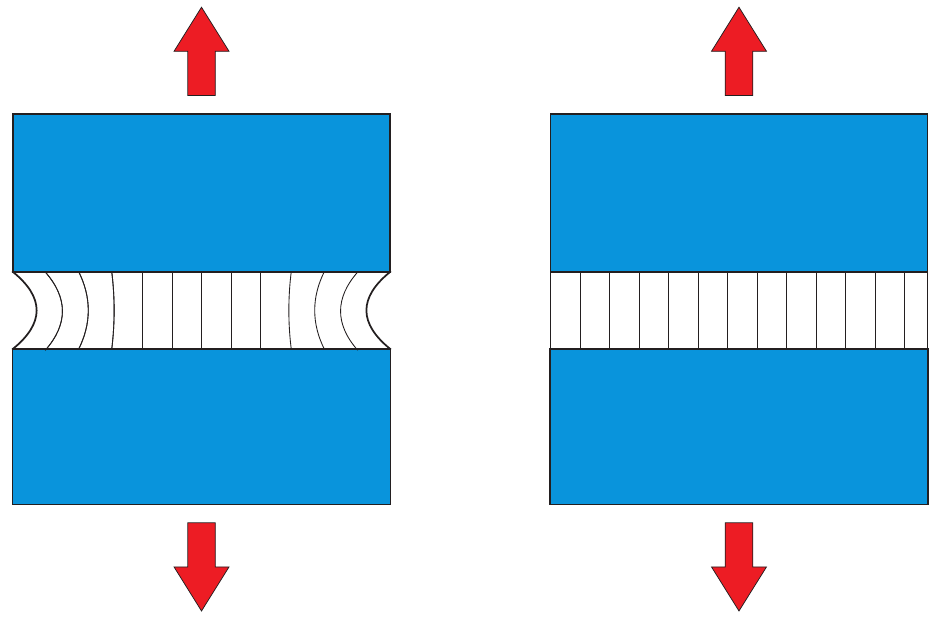}
\caption{Lateral contraction of adhesive joint under tension (left) and lack of lateral contraction in cohesive zone model (right)\label{sketch}}
\end{figure}

Figure \ref{sketch} shows on the left side an adhesive layer under a tensile load. The lateral contraction of the adhesive is partially restricted by the rigid adherents. While there is some lateral contraction visible at the ends of the adhesive layer, there is hardly any lateral contraction possible in its centre. This effect of restricted lateral contraction cannot be modelled directly using a cohesive zone model (right side), because this kind of model does not contain degrees of freedom capable of representing the state of lateral contraction and does not consider any membrane stresses in the adhesive layer. The current article suggests two methods to calculate stiffness parameters for the cohesive zone model in a way that the cohesive zone model gives a good approximation of the correct stiffness of the joint. The required calculation of stiffness parameters requires much less effort than the creation of a detailed model with several elements across the thickness of the adhesive layer.

The stiffness obtained by dividing the elastic modulus by the adhesive layer thickness is a lower bound for the stiffness parameter yielding the correct joint stiffness. An upper bound can easily be derived by assuming complete restriction of lateral contraction and corresponds to an effective modulus $\bar{E}$
\begin{equation}\label{eq:complete}
\bar{E}_{cr} = \frac{1-\nu}{1-\nu-2\nu^2}E
\end{equation}
This effective modulus overestimates the joint stiffness, because it does not consider that the adhesive can contract laterally close to the free surfaces of the adhesive layer. Therefore, the next step in improvement of the accuracy of stiffness prediction is to consider an elastic layer of finite size between rigid adherents.

A first closed-form solution for this kind of elasticity problem was published by Gent and Lindley in 1959 \cite{Gent01061959}. Their purpose was to predict the compression stiffness of rubber blocks, therefore they considered incompressible elastic behaviour. The stiffness was calculated for circular disks and infinite strips. In 1979 Lindley extended the work to compressible elastic materials using an energy method \cite{Lindley01011979}. A "pressure solution" approach was applied by Kelly to calculate the compresion as well as the tilting stiffness \cite{Kelly1997}. A solution for rectangular layers, a shape which is quite often found in adhesive joints, was derived by Koh and Lim \cite{Koh2001445}. The same problem was addressed by Tsai (2005) \cite{Tsai20053395} who published a single-series closed form instead of the double-series solution of Koh and Lim. The single-series solution has the advantage of faster convergence. More recent work considers elastic layers between flexible instead of rigid reinforcements and compression as well as bending and warping loads, e.g. \cite{Pinarbasi2008794}.

In the current paper, the closed-form solution of Tsai for rectangular elastic layers between rigids is used to improve the prediction of adhesive joint normal stiffness in cohesive zone models. Two approaches of different level of complexity are suggested: The first approach simply uses the effective modulus calculated by the solution of Tsai instead of the elastic modulus to calculate the stiffness parameter of the cohesive law. The second approach provides a stiffness depending on the position within the adhesive layer to account for the different severity of restriction of lateral contraction at the edges and in the centre of the elastic layer.

The following section summarizes the solution of Tsai. In section \ref{cohesive} this solution is used to obtain an improved stiffness in cohesive zone models. A numerical validation of this approach by parametric studies using finite elements is shown in section \ref{validation}. Since the modelling effort to create finite element models with locally varying material properties can be considerable, section \ref{automation} deals with the automation of the model creation.

\section{Closed form solution for rectangular layers between rigids}\label{closed}

Tsai \cite{Tsai20053395} analyses a linear elastic layer under a compressive load and derives a solution for its stiffness taking into account the restriction of lateral expansion. Since no non-linearities are considered, the solution applies equally to a tensile load with restricted lateral contraction.

\begin{figure}[htp]
\begin{center}
\includegraphics[width=6cm]{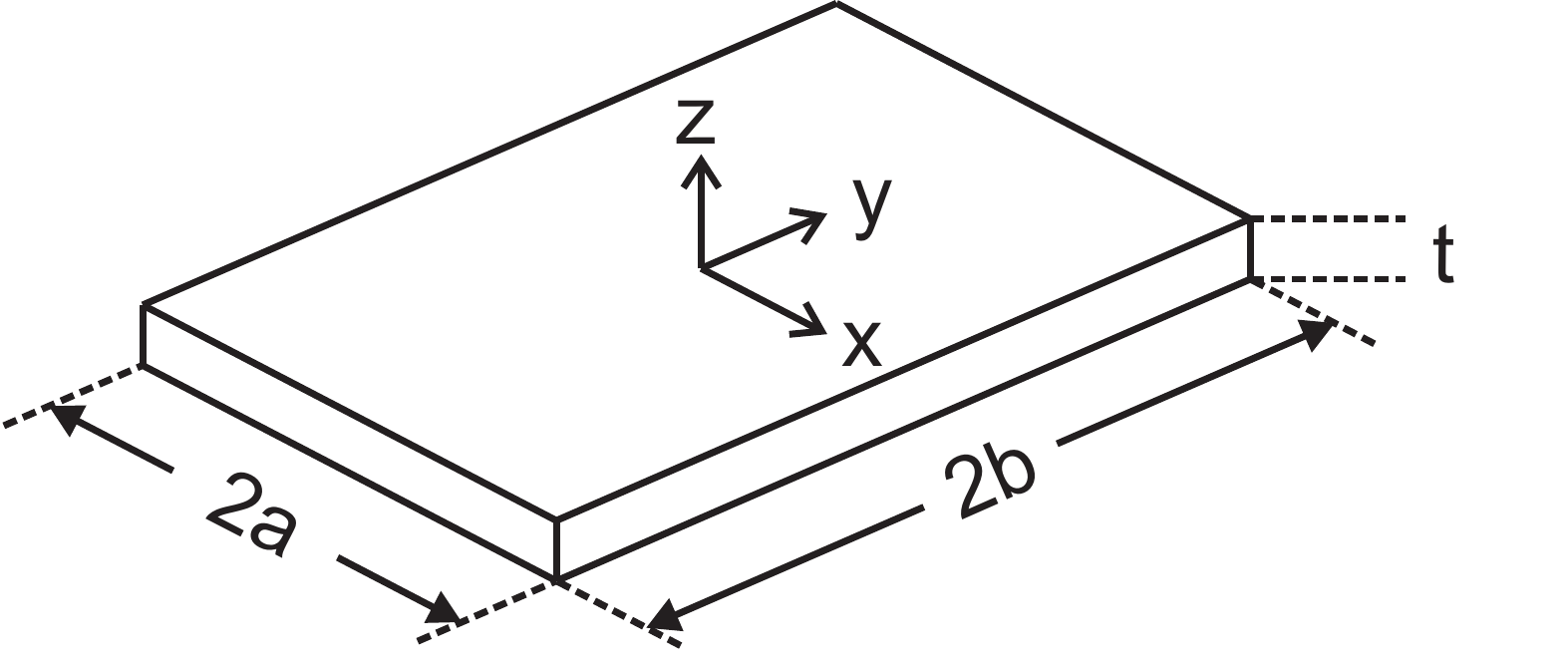}
\end{center}
\caption{Geometry of elastic layer in the model of Tsai (right)\label{fig:tsai}}
\end{figure}

Let us consider a rectangular, isotropic, linear elastic layer between rigid plates (figure \ref{fig:tsai}). A Cartesian coordinate system $(x,y,z)$ is positioned in its centre. The elastic layer has a width of $2a$ along the $x$-axis, a length $2b$ along the $y$-axis and a thickness $t$. Its material behaviour is characterized by the elastic modulus $E$ and the Poisson's ratio $\nu$. According to Tsai \cite{Tsai20053395}, the distribution of normal stress in relation to the applied compressive strain is given by
\begin{equation}\label{eq:sigma}
\frac{\sigma_{zz}(x,y)}{-\varepsilon_c} = \frac{E}{1+\nu} \left(1 + \frac{\nu}{1-2\nu} \frac{p(x,y)}{\kappa\varepsilon_c} \right)
\end{equation}
with the hydrostatic stress distribution calculated by
\begin{align}
\frac{p(x,y)}{\kappa\varepsilon_c} = 1 - &2\nu + 4 \nu \sum_{n=1}^\infty \frac{(-1)^n}{\left(n-\frac{1}{2}\right)\pi} \left\lbrace \vphantom{\frac{\cosh{\beta_n x}}{\cosh{\beta_n a}}}\right. \nonumber\\
&\left[ \left( 1 - \frac{3(1-2\nu)}{\gamma_n^2 t^2 + 6(1-\nu)} \right) \frac{\cosh{\beta_n x}}{\cosh{\beta_n a}} - \frac{\cosh{\gamma_n x}}{\cosh{\gamma_n a}} \right] \cos{\gamma_n y} \nonumber\\
+& \left. \left[ \left( 1 - \frac{3(1-2\nu)}{\bar{\gamma}_n^2 t^2 + 6(1-\nu)} \right) \frac{\cosh{\bar{\beta}_n y}}{\cosh{\bar{\beta}_n b}} - \frac{\cosh{\bar{\gamma}_n y}}{\cosh{\bar{\gamma}_n b}} \right] \cos{\bar{\gamma}_n x} \right\rbrace \label{eq:pressure}
\end{align}
using the abbreviations
\begin{align}
\gamma_n &= \left( n - \frac{1}{2} \frac{\pi}{b} \right) & \bar{\gamma}_n &= \left( n - \frac{1}{2} \frac{\pi}{a} \right) \\
\beta_n &= \sqrt{\gamma_n^2 + \alpha^2} & \bar{\beta}_n &= \sqrt{\bar{\gamma}_n^2 + \alpha^2} \\
\alpha &= \frac{1}{t} \sqrt{\frac{6(1-2\nu)}{1-\nu}} &&
\end{align}

The effective modulus $\bar{E}$ of the layer considering the restriction of lateral expansion is defined by averaging over the adhesive layer
\begin{align}
\frac{\bar{E}}{E} = \frac{\int_{-b}^b\int_{-a}^a \sigma_{zz}(x,y)\,dx\,dy}{-4\,a\,b\,E\,\varepsilon_c} \label{eq:effective}\\
= 1 + \frac{4\nu^2}{(1+\nu)(1-\nu)} \sum_{n=1}^\infty & \frac{1}{\left(n-\frac{1}{2}\right)^2\pi]^2} \left\lbrace \vphantom{\frac{\tanh \gamma_n a}{\gamma_n a}} \right. \nonumber\\
&\frac{\tanh \gamma_n a}{\gamma_n a} - \left[ 1 - \frac{3(1-2\nu)}{\gamma_n^2 t^2+6(1-\nu)} \right] \frac{\tanh \beta_n a}{\beta_n a} +  \nonumber\\
& \left. \frac{\tanh \bar{\gamma}_n b}{\bar{\gamma}_n b} - \left[ 1 - \frac{3(1-2\nu)}{\bar{\gamma}_n^2 t^2+6(1-\nu)} \right] \frac{\tanh \bar{\beta}_n b}{\bar{\beta}_n b} \right\rbrace \label{eq:tsai}
\end{align}

These results of Tsai will be applied in the next section to cohesive zone modelling. Tensile loads instead of compressive loads will be considered in the following sections, but this switches only the sign of the compressive strain $\varepsilon_c$ without altering the solution of Tsai.

\section{Cohesive zone stiffness for rectangular adhesive layers}\label{cohesive}

In the current section two approaches to improve the accuracy of normal stiffness in the cohesive zone model of a rectangular adhesive layer are proposed. The first approach is rather straightforward. The mode I stiffness parameter $k_I$ for entire the adhesive layer is calculated from the effective modulus \eqref{eq:tsai} and the layer stiffness:
\begin{equation}\label{eq:k1simple}
k_I = \frac{\bar{E}}{t}
\end{equation}
If this approach is applied to finite element modelling, it simply means that one material parameter of the cohesive law needs to be adjusted, but no additional modelling effort is involved. Since the series in \eqref{eq:tsai} converges rather fast, the calculation of the effective modulus imposes no difficulties.

In the case of homogeneous strain, i.e.~an adhesive layer between rigid substrates, this method yields very accurate results for the joint stiffness. However, in other cases the approach can differ from the exact solution, because it neglects one important feature of the lateral constraint effect: its inhomogeneity. Close to the free surfaces of the adhesive layer, there is a significant amount of lateral contraction possible, while it is severely constrained in the centre of the layer. To account for this effect, the second approach suggested here defines the stiffness dependent on the location within the adhesive layer.

Again, the closed-form solution of Tsai for rigid substrates is used. The stiffness of the cohesive law is chosen according to the stress distribution \eqref{eq:sigma} in the closed form solution:
\begin{equation}\label{eq:k1local}
k_I(x,y) = \frac{1}{t} \frac{\sigma_{zz}(x,y)}{-\varepsilon_c}
\end{equation}
In the case of rigid substrates this yields the same joint stiffness as the simple approach \eqref{eq:k1simple} due to the definition of the effective modulus \eqref{eq:effective}.

In a finite element model the locally varying stiffness \eqref{eq:k1local} is optimally prescribed for each integration point. Alternatively, it can be set for each element at a certain loss of accuracy. It should be noted that due to discretization the total stiffness of the layer between rigid substrates can differ from the solution \eqref{eq:k1simple}. This discretization error can be compensated by multiplying the stiffness of all elements by a correction factor.

The shear stiffness in the cohesive law is unchanged for both approaches and simply determined by the shear modulus:
\begin{equation}
k_{II} = \frac{G}{t} = \frac{E}{2(1-\nu)t}
\end{equation}

\section{Validation}\label{validation}

After proposing two methods to improve the normal stiffness calculated by cohesive zone models of rectangular adhesive layers, the methods will be tested in extensive parametric numerical studies. For each parameter set in these studies, five finite element solutions will be compared:
\begin{itemize}
\item A reference solution is generated by using small solid elements to discretize the adhesive layer, denoted by "R".
\item A cohesive zone model is used where the normal stiffness is derived from the elastic modulus ($k_I=E/t$), denoted by "E".
\item A cohesive zone model is used where the normal stiffness is derived from modulus considering complete lateral constraint \eqref{eq:complete}, denoted by "C".
\item A cohesive zone model is used where the normal stiffness is derived from the effective modulus of Tsai \eqref{eq:k1simple}, denoted by "T".
\item A cohesive zone model is used where the normal stiffness is spatially varying according to \eqref{eq:k1local}, denoted by "V".
\end{itemize}
The adhesive layer was discretized by a single layer of cohesive elements. The in-plane edge length of the cohesive elements was chosen to be one quarter of the adhesive layer thickness. The solid elements of the reference models were chosen of approximately cubical shape with 10 elements across the adhesive layer thickness. In case that the number of elements across the length or width of the layer exceeded 200, the element length in that direction was increased. This restriction of numerical effort was necessary to perform the 3375 simulations of the parametric studies within a reasonable amount of time.

The numerical effort was reduced further by performing a linear analysis only. This restriction also simplifies the comparison of the solutions of the different models, since the response of the joint can be expressed by its stiffness with no need to compare non-linear force-displacement curves.

Since the modelling of the initial stiffness of the joint and not the fracture process is the topic of this validation, the damage behaviour of the cohesive zone model does  has no impact on the results of the validation. They are applicable to all cohesive laws which exhibit an uncoupled linear behaviour prior to damage initiation, e.g. bilinear or trapezoidal traction-separation relations.

\subsection{Rigid substrates}

The first parametric study regards the simple case of tension applied to an adhesive layer between rigid substrates. In this case the joint stiffness is proportional to the elastic modulus. At constant aspect ratios it is inversely proportional to the layer thickness. Therefore, only three parameters need to be considered in this parametric study. The parameters and their considered values are given in table \ref{tab:para1}.
\begin{table}
\setlength{\extrarowheight}{3.5pt}
\caption{Parameters of parametric study considering rigid adherents\label{tab:para1}}
\begin{center}
\begin{tabular}{c|c|c|c|c|c}
Parameter & Value 1 & Value 2 & Value 3 & Value 4 & Value 5 \\\hline
$a/t$ & 5 & 10 & 20 & 40 & 80 \\\hline
$b/t$ & 5 & 10 & 20 & 40 & 80 \\\hline
$\nu$ & 0.3 & 0.45 & 0.49 & &
\end{tabular}
\end{center}
\end{table}

For each simulation the stiffness $S=F/u$ was calculated and compared with the stiffness $S_{ref}$ of the reference model using a fine solid mesh. A ratio of stiffness was considered to quantify the deviation $f$ from the reference solution:
\begin{equation}
f = \left\lbrace
\begin{aligned} \frac{S}{S_{ref}} \quad & \text{if } S > S_{ref} \\
\frac{S_{ref}}{S} \quad & \text{if } S \le S_{ref} \end{aligned}
\right.
\end{equation}

In case of the cohesive zone model simply using the elastic modulus, this deviation $f$ is in the interval between 1.29 and 17. The deviation can be significantly reduced to the interval between 1.00 and 2.3 by using the completely constrained modulus. In 44\% of the test cases the deviation is still larger than 1.1, in 22\% larger than 1.2.

The cohesive zone models using the effective modulus of Tsai or employing the spatially varying modulus agree with the reference solution within the accuracy limits of this numerical study ($f<1.006$). This was expected because the test cases of this parametric study fulfill the assumptions of the closed form solution.

\subsection{Flexible substrates}

\begin{figure}[htp]
\begin{center}
\includegraphics[width=6cm]{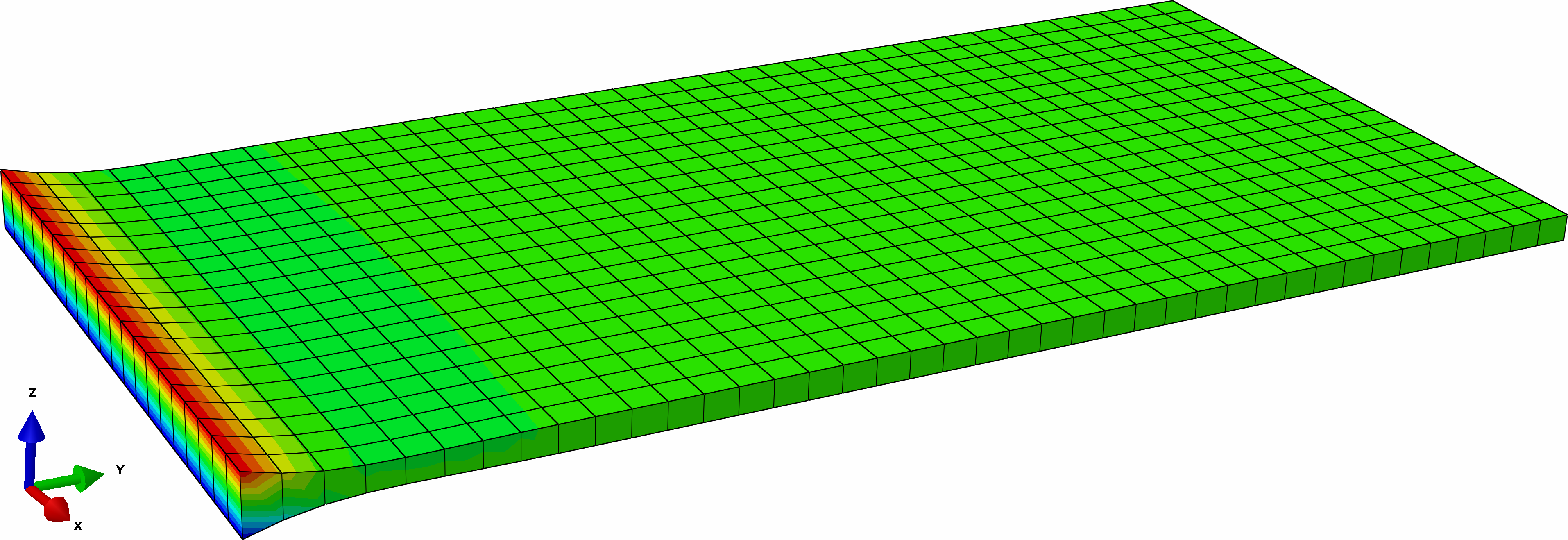} \hspace{1cm}
\includegraphics[width=4.5cm]{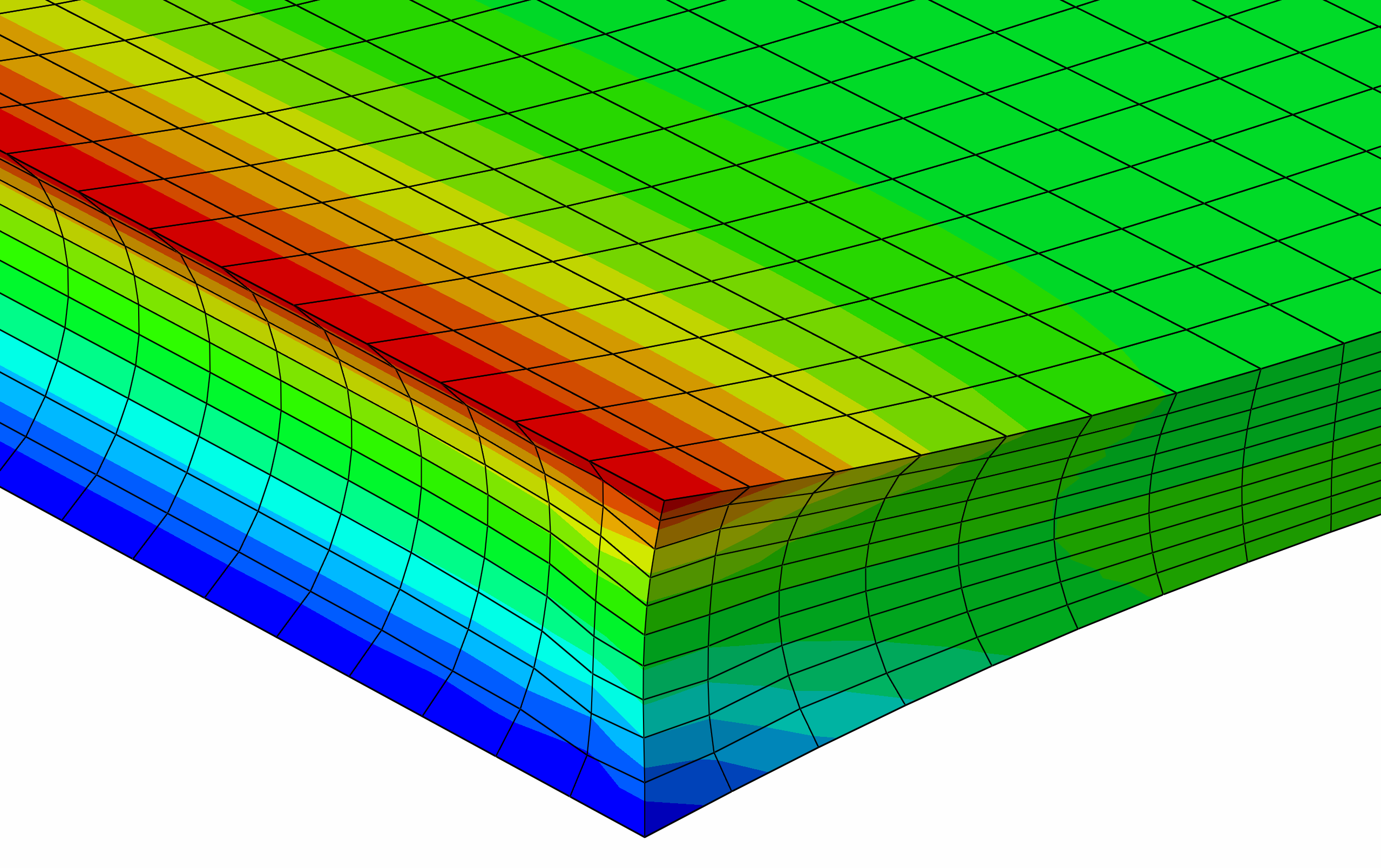}
\end{center}
\caption{Finite element models for second parametric study: cohesive zone model (left) and detail of reference model (right)\label{fig:para2}}
\end{figure}

The second parametric study considers a less simple and more realistic test case. The adherents are no longer rigid but have the elastic properties of steel sheets ($E = 210$ GPa, $\nu = 0.3$). In the finite element models they are represented by shell elements. The shell elements shared nodes with the cohesive or solid elements of the adhesive layer. An example of one of the cohesive zone models is displayed on the left side of figure \ref{fig:para2}. The colours in this contour plot correspond to the displacements in thickness direction of the adhesive layer. Obviously, different from the first parametric study no homogeneous load is applied. Instead, the element motion is prescribed only for the nodes along the left edge of the adhesive layer. The left edge of the bottom adherent is fixed while the corresponding edge of the upper adherent is moved normal to the adhesive layer. A rotation around the edge is allowed for both edges. This way the adhesive layer is subjected to a kind of peel load similar to a T-peel test.

Since the deformation of the joint depends on the relative stiffness of adhesive and adherents, a larger number of parameters needs to be considered than in the first parametric study. For each of the 5 modelling approaches 600 finite element simulations were performed using the parameter values given in table \ref{tab:para2}. The adherent thickness is denoted by $t_a$.
\begin{table}
\setlength{\extrarowheight}{3.5pt}
\caption{Parameters of parametric study considering flexible adherents\label{tab:para2}}
\begin{center}
\begin{tabular}{c|c|c|c|c|c|c}
Parameter & Unit & Value 1 & Value 2 & Value 3 & Value 4 & Value 5 \\\hline
$a$ & mm & 5 & 10 & 20 & 40 & 80 \\\hline
$b$ & mm & 5 & 10 & 20 & 40 & 80 \\\hline
$t$ & mm & 0.3 & 2 &  &  &  \\\hline
$\nu$ & - & 0.3 & 0.45 & 0.49 & & \\\hline
$E$ & MPa & 10 & 2000 &  &  &  \\\hline
$t_a$ & mm & 0.8 & 1.5 &  &  &
\end{tabular}
\end{center}
\end{table}

\begin{figure}[p]
\centering
\includegraphics[width=6.5cm]{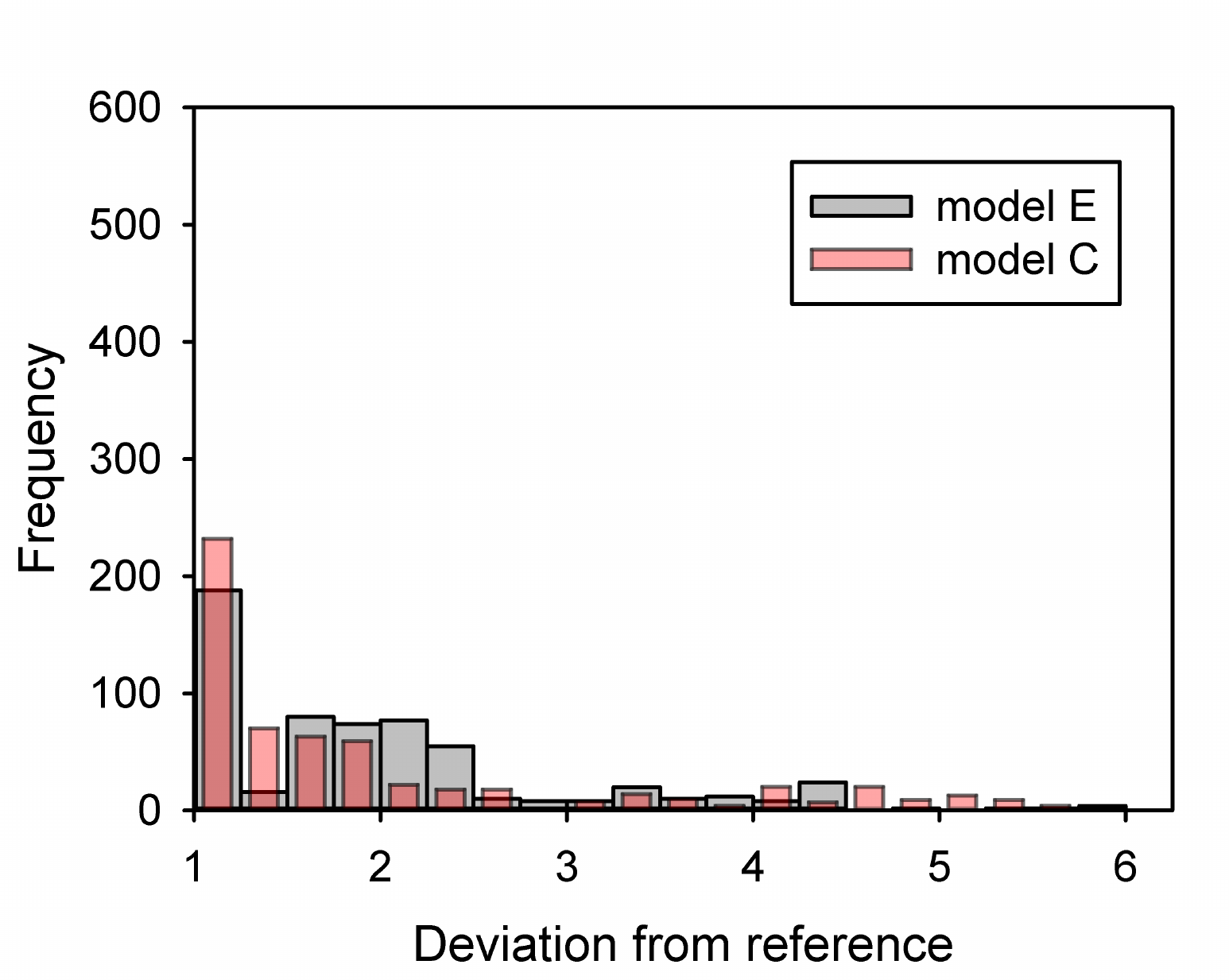} \\
\includegraphics[width=6.5cm]{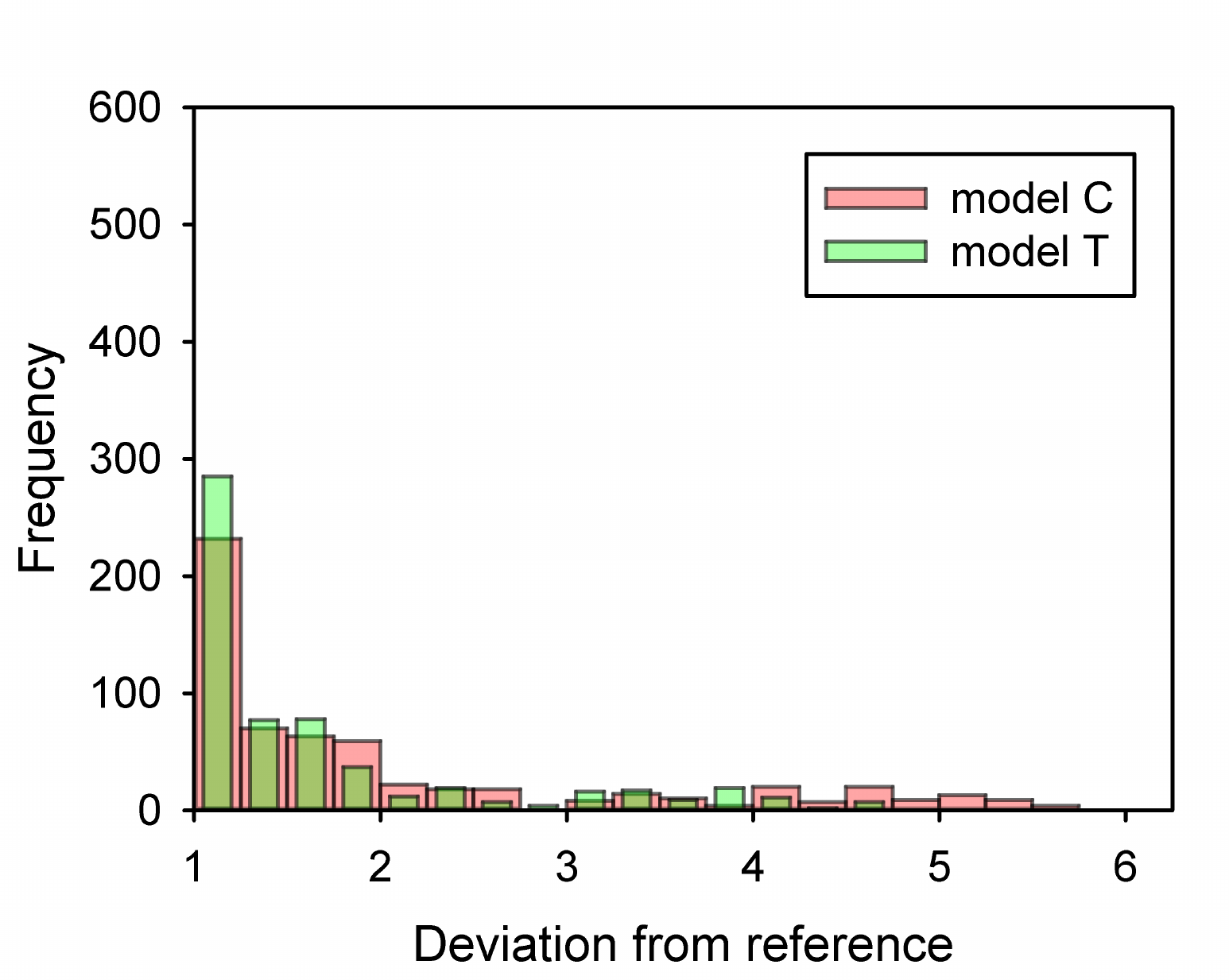} \\
\includegraphics[width=6.5cm]{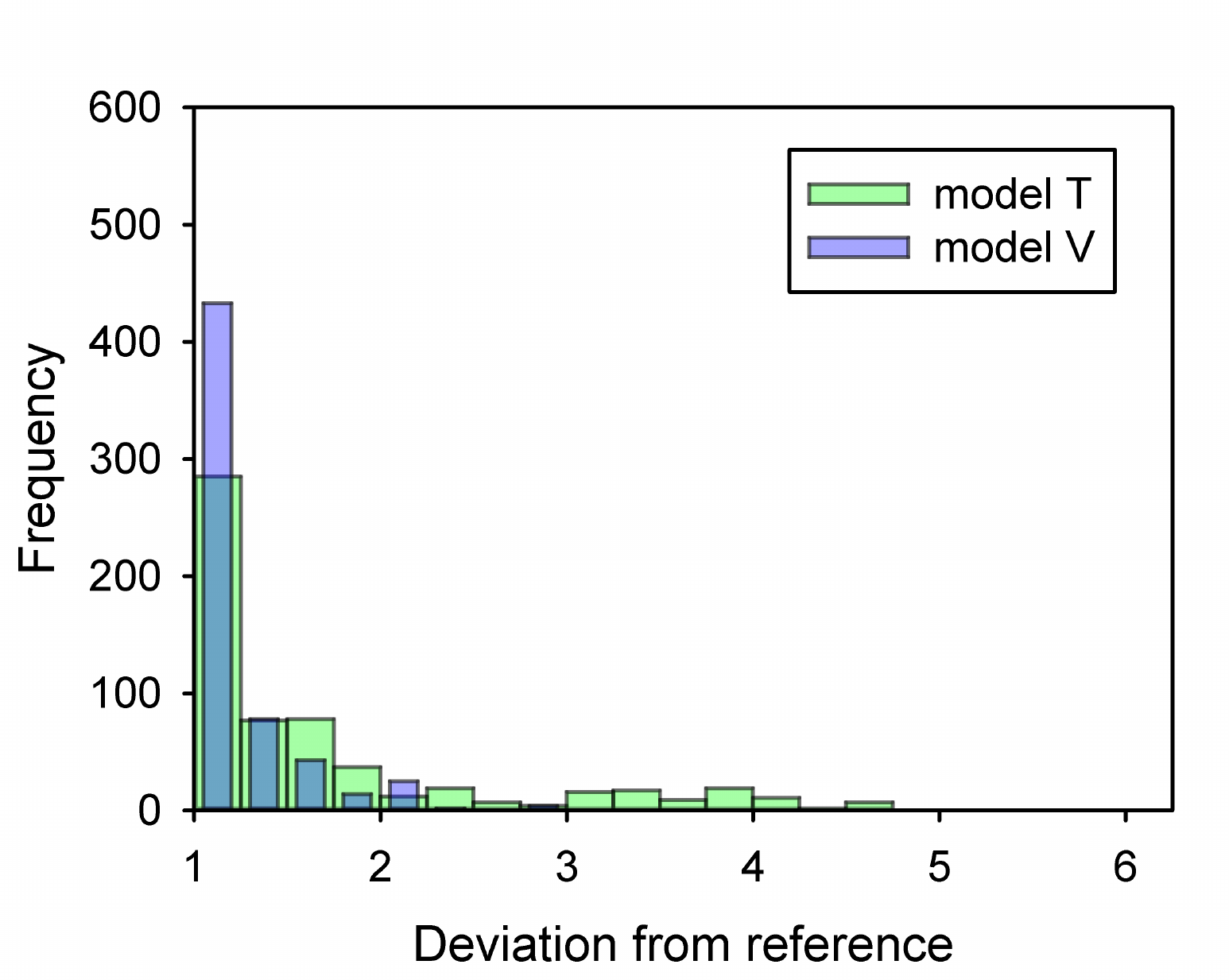} 
\caption{Histograms of deviation from reference solution}
\label{fig:histo}
\end{figure}

The deviation of the stiffness of the cohesive zone models from the reference solutions was calculated like in the first parametric study. Figure \ref{fig:histo} shows histograms of the deviation. The cohesive zone model "E" using directly the elastic modulus shows a deviation below 1.25 in slightly less than a third of the 600 test cases. A large part of the tests resulted in deviations up to 2.5, and a small part differed from their reference solution up to a factor 6. By using the completely constrained modulus instead of the elastic modulus (model "C"), the fraction of tests with small deviations from the reference can be increased. On the other hand, the number of test cases leading to very large deviations increases as well. Neither of these approaches allows a very reliable prediction of normal stiffness.

The second histogram shows that a significant improvement is gained if the effective modulus of Tsai is used (model "T") instead of the elastic modulus or the completely constrained modulus. Nearly one half of the tests yielded deviations smaller than 1.25, and the number of tests with large deviations decreased. The best results are obtained using the spatially varying modulus (model "V"). More than two thirds of the tests are now covered by the bar of the smallest deviation, and the width of the distribution has become much smaller.

An alternative way to evaluate the results is to look for the model giving the best stiffness prediction for every test case, respectively. In 63\% of the test cases the prediction of the model "C" was better than model "E". Model "T" was better than model "E" in 75\% and better than model "C" in all test cases. The spatially varying modulus, model "V", resulted in the best stiffness prediction in 85\% of the tests. In a third of the remaining 15\% the deviation was quite small anyhow ($f<1.1$), so that only in 10\% of the tests the model "V" was not the optimal or a very good choice. These 10\% consisted of the following parameter sets:
\begin{itemize}
\item stiff adhesive layer ($E=2000$ MPa, $t=0.3$ mm), nearly incompressible ($\nu=0.49$), thick adherent ($t_a=1.5$ mm): This case is of little interest, because adhesives with a high elastic modulus usually have smaller Poisson ratios.
\item small bonding surface ($a=b=5$ mm), nearly incompressible ($\nu=0.49$), thin adhesive ($t=0.3$ mm) and thick adherent ($t_a=1.5$ mm): In these two cases the model "E" was slightly better than model "V".
\item flexible elastic layer ($E=10$ MPa, $t=2$ mm), thin adherent ($t_a=0.8$ mm), $\nu=0.45$: In these 4\% of the test cases model "T" provided better stiffness predictions.
\item flexible elastic layer ($E=10$ MPa, $t=2$ mm), thin adherent ($t_a=0.8$ mm), nearly incompressible ($\nu=0.49$): In about a half of these 4\% of the test cases model "T" provided better stiffness predictions, especially for small bonding surfaces.
\end{itemize}

After comparing the accuracy of the different models, now the effect of each parameter on the quality of stiffness prediction shall be considered for each model.
\begin{itemize}
\item An increase in Poisson's ratio $v$ causes an increase of the deviation from the reference solution. Only for the model "V" there are 12\% exceptions with thick adherents and mostly thick adhesive layers.
\item A thick adherent ($t_a=1.5$ mm) improves the prediction of model "E" but increases the error of models "C" and "T" in 90\% of the test cases. Most of the exceptions have a small Poisson's ratio ($\nu=0.3$), high modulus and thick adhesive layer. The results of model "V" are mixed: The prediction for thin adhesive layers benefits from thick adherents while thick adhesive layers are modelled more accurately in case of thin adherents.
\item The influence of the adhesive layer thickness $t$ is unambiguous for model "C" where the error of a model with a thin layer ($t=0.3$ mm) is always smaller than of the same model with a thick ($t=2$ mm) layer. This relation is reversed for model "E" in 90\% of the cases, exceptions exhibit a small adhesive modulus, large Poisson's ratios and small bonding surfaces. Both models "T" and "V" show a mixed result with \sfrac{2}{3} of the test cases benefiting from thick adhesive layers. While the number of cases with the same influence of the adhesive thickness is the same for models "T" and "V", the interaction with the other parameters is not. Using model "T" we get an increase of the error with increasing thickness in cases of a stiff adhesive where the Poisson's ratio is small or the length $a$ is large. Model "V" shows the same tendency for thin adherents and small bonding surfaces or small Poisson's ratios.
\item A high adhesive modulus ($E=2000$ MPa) results in a higher accuracy of model "E" than a low modulus ($E=10$ MPa). Models "C" and "T" show the opposite behaviour with only 3\% exceptions. The accuracy of model "V" is better for a high modulus in \sfrac{2}{3} of the test cases.
\item The parameters $a$ and $b$ describing the size of the bonding surface have a different influence on the accuracy of predictions, because the loading in this parametric study is not symmetric with respect to a switch of coordinate directions $x$ and $y$. $a$ is the length of the edge of the adhesive layer where the load is applied while $b$ is the width in the perpendicular direction. An increase of length $a$ results in an increase of the error of model "E" but a decrease of the error of model "C". The influence of $a$ on the accuracy of model "T" depends on the adhesive layer thickness: Models with thin layers most often gain accuracy with increasing $a$ while the influence is reversed for the thick layer. The model "V" benefits from increasing $a$ in \sfrac{5}{6} of the test cases.
\item While the prediction accuracy was either monotonously increasing or decreasing with increasing parameter $a$ in nearly all cases, the effect of parameter $b$ is in general not that simple.
\end{itemize}

Apart from the possibility to compare the accuracies of the models and to study the influence of parameters on it, the parametric study also allows to indicate the current limitations of stiffness predictions using cohesive zone models. For this aim we shall consider the results of the most accurate of the models, which is the model "V" using the spatially varying modulus. In several cases no deviation of the predicted stiffness from the reference solution exceeding the numerical errors of the reference solution was observed, in the worst case the prediction differed from the reference by a factor of 3.2. All test cases resulting in deviations in the range from 1.82 to 3.2 have in common that the adhesive layer is thin ($t=0.3$ mm), the Poisson's ratio high ($\nu=0.49$), and the adherent is thick ($t_a=1.5$ mm). Since nearly incompressible adhesives are most often applied in flexible joints with thicker adhesive layers, these cases are of little concern to the practical application. The next range, between deviation 1.54 and 1.82, still consists only of cases with the high Poisson's ratio high ($\nu=0.49$). The first model with a lower Poisson's ratio ($\nu=0.45$) is found at a deviation of 1.54, but this model contains the extreme case of a small adhesive layer with $a=b=5$ mm. If we consider the test cases with both $a$ and $b$ at least 10 mm, than the worst model with $\nu=0.45$ deviates from the reference stiffness only by a factor 1.35.

In summary it can be concluded that the use of the effective modulus instead of simply the elastic modulus or the completely laterally constrained modulus results in an improvement of joint stiffness prediction. A larger improvement can be obtained using the method of the spatially varying modulus. While these methods provided good results on a broad range of test cases, their predictions for thin layers of low modulus adhesives still deviate significantly from the reference solutions.

\subsection{Mesh dependence}

The size of cohesive elements chosen for the parametric studies was quite small compared to many applications of cohesive elements in industrial practice. Therefore, it was possible to validate the approaches to consider the lateral contraction constraint in cohesive zone models without much disturbation by discretization errors.

This leaves the question if the increase in accuracy can also be gained by the two suggested methods if coarse finite element meshes are used. Two test cases from the parametric study with flexible adherents have been chosen to study mesh size effects, see table \ref{tab:para3}.
\begin{table}[htb]
\setlength{\extrarowheight}{3.5pt}
\caption{Parameter sets of discretization studies\label{tab:para3}}
\begin{center}
\begin{tabular}{c|c|c|c|c|c|c}
Test case & $a$ & $b$ & $t$ & $E$ & $\nu$ & $t_a$ \\\hline
A & 20 mm & 40 mm & 0.3 mm & 2 GPa & 0.45 & 0.8 mm \\\hline
B & 20 mm & 20 mm & 2 mm & 10 MPa & 0.49 & 1.5 mm
\end{tabular}
\end{center}
\end{table}

\begin{figure}[htb]
\centering
\includegraphics[width=6cm]{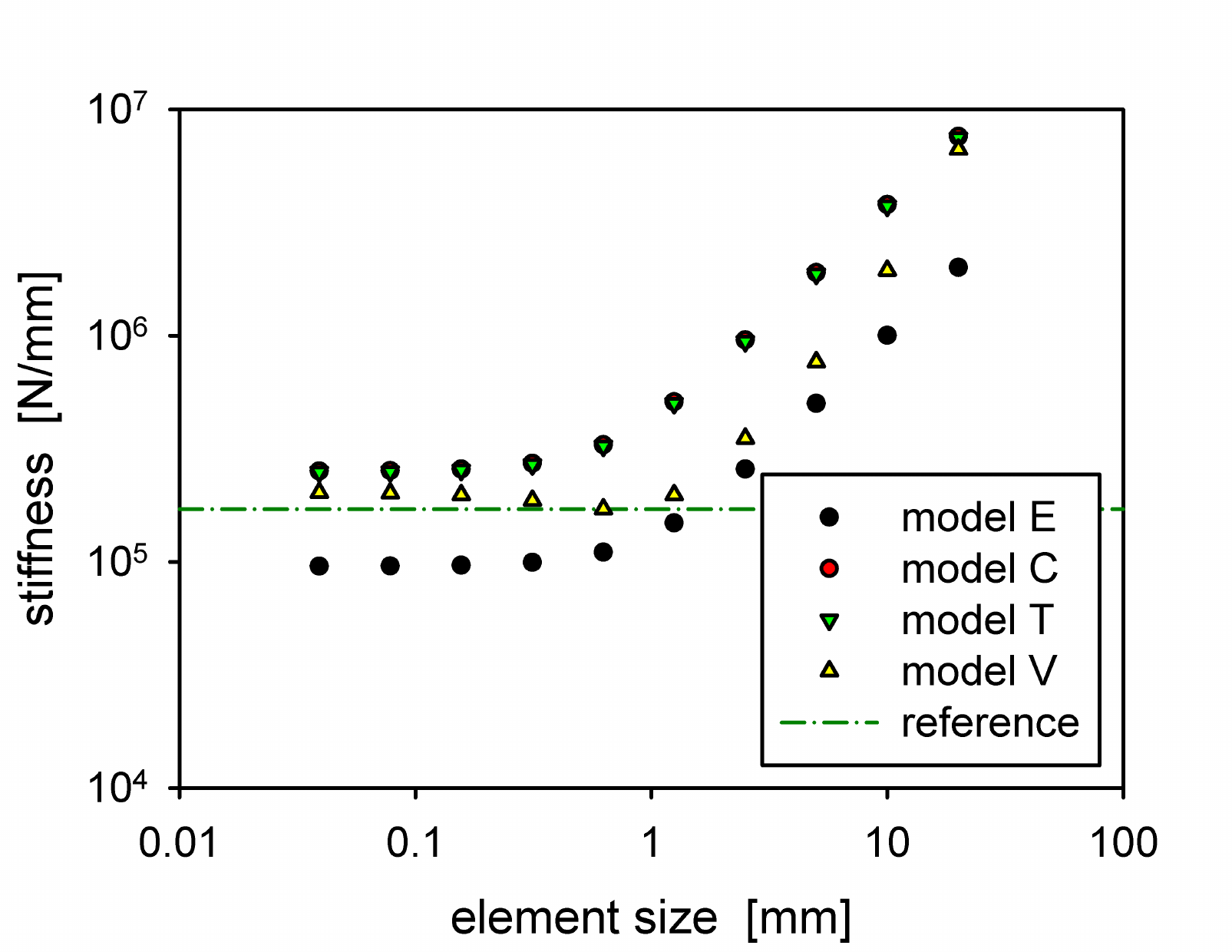} \hfill
\includegraphics[width=6cm]{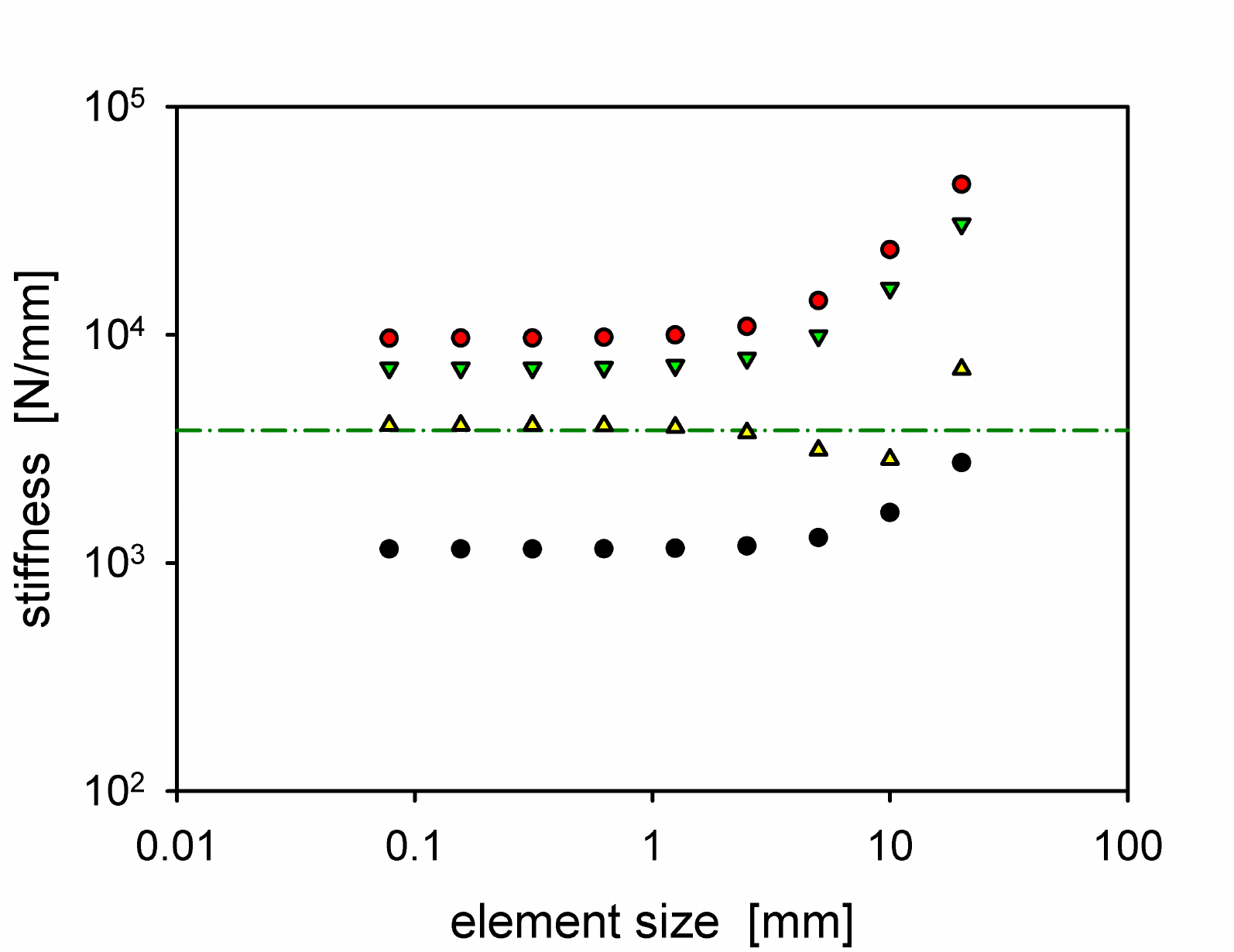}
\caption{Mesh dependence of stiffness in test case A (left) and B (right)}
\label{fig:discr}
\end{figure}

Figure \ref{fig:discr} shows the stiffness obtained from models using different element sizes for the cohesive elements. Convergence for small elements can be seen as well as large mesh effects if only two or four elements span the width of the layer. Both the local bending of the adherents as well as the inhomogeneity of the lateral contraction constraint can not be represented in the finite element model if the elements are too large. Therefore, the stiffness is overestimated by coarse meshes. In test case A the model using the spatially varying modulus reached a good stiffness prediction at an element size of 1.25 mm, in test case B at 2.5 mm.

\section{Automation}\label{automation}

While a finite element model of an adhesively bonded structure containing a traction-separation law for the cohesive elements can be improved quite easily by modifying the stiffness parameter according to the effective modulus \eqref{eq:k1simple}, the use of a spatially varying stiffness in the adhesive layer requires substantial modelling effort. Therefore, the second method remains of purely academical interest and not suited to practical application, unless this additional modelling effort can be removed by automation.

To show the feasibility of such an automation, a Python script was developed to apply the spatially varying modulus in finite element models using the FE software Abaqus. The script is supposed to read an existing model containing an adhesive layer modelled by a classical traction-separation law, and to create the necessary changes to consider the effect of constrained lateral contraction.

As an input the script needs the elastic constants $E,\nu$ of the adhesive, and its damage initiation and failure definitions which will remain unchanged. Furthermore, it reads the output data base of the existing model to get the nodal coordinates of the adhesive layer. A node set name must be supplied to identify the adhesive layer in the model.

The adhesive layer must be rectangular, but may be arbitrarily positioned and oriented. The script determines the geometry of the layer and calculates the local stiffness at each node. An "initial field" is used to store the information on the spatial stiffness distribution, a field dependent elastic behaviour is used to assign the stiffness to the cohesive elements.

Two files are created by the script and can be included by the input file of the finite element model. One file contains the initial field definitions, the other file replaces the adhesive material definition of the original model.

\section{Conclusions}\label{conclusions}

The primary intention of cohesive zone models of adhesive joints is not an accurate joint stiffness prediction but the simulation of the entire damage process at a low numerical cost. Cohesive zone models lack the capability to describe the lateral contraction caused by tensile loads and its constraint by the adherents. Still, a cohesive zone model can provide the correct joint stiffness if the stiffness of the cohesive zone model is adjusted accordingly.

Two methods to calculate an appropriate stiffness parameter for the traction-separation law for rectangular adhesive layers have been suggested in this paper. Both of them use a closed-form solution for rectangular elastic layers between rigid plates under compression load developed by Tsai \cite{Tsai20053395}.

The first approach calculates an effective stiffness depending on the adhesive layer geometry and its elastic constants \eqref{eq:tsai}. This effective modulus divided by the layer thickness is then used as normal stiffness of the cohesive zone model. No additional modelling effort is required. A large parametric study showed that this method improves the joint stiffness prediction significantly compared to the conventional choice of the elastic modulus or the completely constrained elastic modulus for the cohesive law.

The second approach considers that the lateral contraction of the adhesive layer is less constrained close to its free surfaces than in its centre. Consequently, the method generates a spatially varying stiffness. It gains a further increase in accuracy of the joint stiffness compared to the first approach. Its application in a finite element simulation increases the modelling effort, but most of the additional effort can be removed by automation.

The spatially varying stiffness has its largest gradients close to the free surfaces of the adhesive layer. Therefore, a sufficiently fine mesh is necessary to consider its effect in a finite element model. The investigation of mesh size effects in two test cases showed that element sizes of several millimetres as are often used in automotive crash simulations today may not always be sufficient to calculate the joint stiffness under peel loads.

So far only rectangular adhesive layers have been considered. Since similar closed-form solutions exist for circular layers, the methods can easily be transferred to circular adhesive layers. More arbitrarily shaped adhesive layers require another approach. Perhaps it is possible to assign a spatially varying stiffness based on the distance from the free surface of the adhesive layer, but this will probably lead to a less accurate stiffness prediction than for the simple geometries.

It should be noted that not all simulations involving cohesive zone models require a good prediction of stiffness, and not all models targeting at the structural stiffness require an accurate modelling of the joint stiffness. For example, the compliance of structural adhesive joints in automotive applications is often negligible compared to the compliance of the bonded metal sheets.

\section{Acknowledgement}

The IGF project 17276 N "Bruchverhalten von Klebverbindungen und Koh{\"a\-}siv\-zonenmodell -- Einfluss der Herstellung und Alterung" of the research association Forschungsvereinigung Stahlanwendung e.V.~(FOSTA), Sohnstra\ss{}e 65, D-40237 D{\"u}sseldorf was funded by the AiF under the program for the promotion of joint industrial research and development (IGF) by the Federal Ministry of Economics and Energy based on a decision of the German Bundestag.

\bibliography{paper}

\end{document}